\documentclass[letterpaper, 10 pt, conference]{ieeeconf}  

\IEEEoverridecommandlockouts                              

\usepackage{graphics} 
\usepackage{epsfig} 
\usepackage{mathptmx} 
\usepackage{times} 
\usepackage{amsmath} 
\usepackage{amssymb}  
\usepackage{float} 
\usepackage{subfigure}
\usepackage{multirow}
\usepackage{enumerate}
\usepackage[dvipsnames]{xcolor}
\definecolor{Jinglun}{RGB}{25, 230, 25}

\title{\LARGE \bf
GPR-based Subsurface Object Detection and Reconstruction Using Random Motion and DepthNet}

\author{Jinglun Feng$^{1, \dagger}$, Liang Yang$^{1, \dagger}$, Haiyan Wang$^{1}$, Yifeng Song$^{2}$, Jizhong Xiao*$^{1}$
\thanks{$^{1}$ Electrical Engineering Department, The City College of New York, New York, USA. $\dagger$ The first two authors are equally contributed.
        {\tt\small jfeng1,lyang1@ccny.cuny.edu}, {\tt\small hwang005@citymail.cuny.edu}, the corresponding author is {\tt\small jxiao@ccny.cuny.edu}}%
\thanks{$^{2}$University of Chinese Academy of Sciences,Shenyang Institute of Automation, Chinese Academy of Sciences
        {\tt\small songyifeng@sia.cn}}%
}

\begin{document}

\maketitle
\thispagestyle{empty}
\pagestyle{empty}

\begin{abstract}
Ground Penetrating Radar (GPR) is one of the most important non-destructive evaluation (NDE) devices to detect the subsurface objects (i.e. rebars, utility pipes) and reveal the underground scene. One of the biggest challenges in GPR based inspection is the subsurface targets reconstruction. In order to address this issue, this paper presents a 3D GPR migration and dielectric prediction system to detect and reconstruct underground targets. This system is composed of three modules: 1) visual inertial fusion (VIF) module to generate the pose information of GPR device,  2) deep neural network module (i.e., DepthNet) which detects B-scan of GPR image, extracts hyperbola features to remove the noise in B-scan data and predicts dielectric to determine the depth of the objects, 3) 3D GPR migration module which synchronizes the pose information with GPR scan data processed by DepthNet to reconstruct and visualize the 3D underground targets. Our proposed DepthNet processes the GPR data by removing the noise in B-scan image as well as predicting depth of subsurface objects. For DepthNet model training and testing, we collect the real GPR data in the concrete test pit at Geophysical Survey System Inc. (GSSI) and create the synthetic GPR data by using gprMax3.0 simulator. The dataset we create includes $350$ labeled GPR images. The DepthNet achieves an average accuracy of $92.64\%$ for B-scan feature detection and an $0.112$ average error for underground target depth prediction. In addition, the experimental results verify that our proposed method improve the migration accuracy and performance in generating 3D GPR image compared with the traditional migration methods. 
\end{abstract}

\section{Introduction}

Ground Penetrating Radar (GPR) has become an important tool for subsurface non-destructive inspection \cite{hugenschmidt2006gpr}. By using a GPR cart, subsurface inspection on bridge decks and other concrete structures becomes a routine task in addition to visual inspection of surface defects  \cite{les1994quantitative,daniels1996surface,benedetto2012gpr,benedetto2012novel}. However, current GPR inspection still relies on on-site engineers to push the GPR cart along the survey grid lines to collect GPR data. Furthermore, the conventional B-scan data is difficult to interpret and requires experienced geophysicist to reveal the underground objects. It is desirable to design a new GPR system and migration algorithms to automatically collect GPR data in random motion pattern and analyze the data to reconstruct the underground objects.

\begin{figure}[h]
    \centering
    \includegraphics[width=0.42\textwidth]{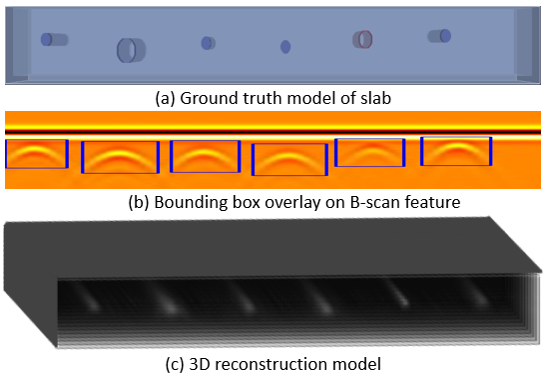}
    \caption{ 3D GPR migration and depth prediction is an challenging problem. We aim to provide a systematic approach to reconstruct the subsurface objects.}
    \label{fig:proposed_system}
\end{figure}


\begin{figure*}[ht]
    \centering
    \includegraphics[width=1.0\textwidth]{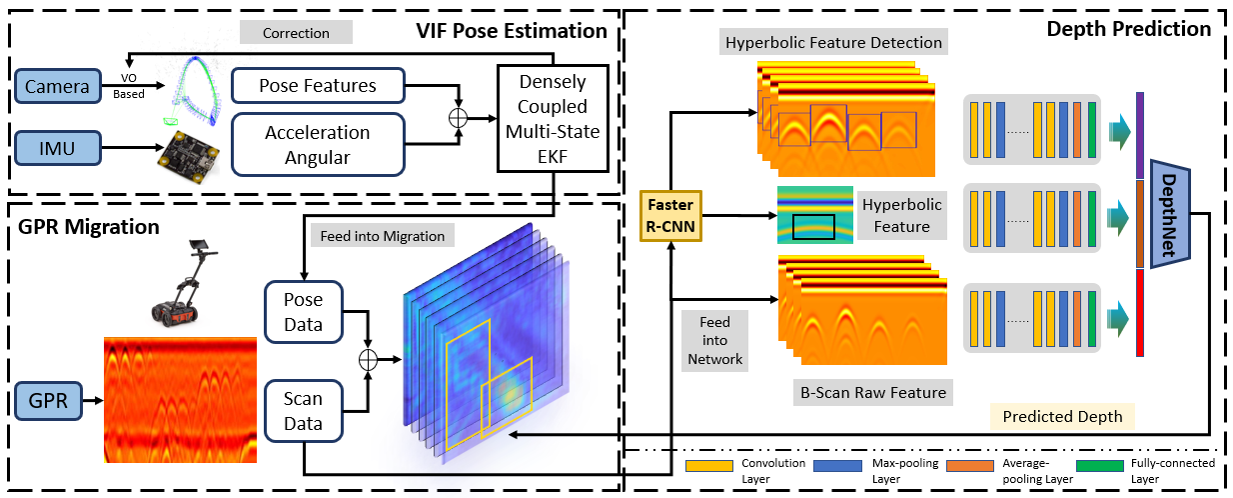}
    \caption{Flow chart of the proposed DNN based GPR Migration framework. The whole system consists of three modules: visual inertial odometry positioning system, GPR migration system and DepthNet neural network model for 3D subsurface targets reconstruction.}
    \label{fig:schmatic}
\end{figure*}


When a GPR system is used for utility survey in outdoor inspection \cite{blindow2010geometry,urbini2001gpr}, GPS is available to provide pose information \cite{yoder2001mapping,ferrara2018gpr}. However, for indoor inspection, we still need to find a way to obtain positioning information since GPS could not be used in indoor environment. Besides, the current commercial GPR cart must move along either horizontal or perpendicular lines of a pre-defined survey grid to trigger GPR scan by survey wheel to recover the underground objects. 


In order to generate 3D structure from the B-scan GPR data, migration algorithm  \cite{bradford2012gpr,leuschen2001matched,feng2004pre} is the most important step to achieve this goal. Authors in \cite{ristow1994fourier}  proposed a hybrid migration method which is Fourier finite-difference migration, it achieves the complex underground targets reconstruction. A full-resolution GPR imaging method is proposed in \cite{grasmueck2005full}, by obtaining a spatial sampling of GPR recording, noninvasive GPR imaging could be generated. Moreover, \cite{qu2012investigation} introduces a migration imaging method for stepped frequency continuous wave (SFCW) GPR system, which is based on compressive sensing algorithm. With this approach, the delivery of high-quality GPR image of underground region is prominent and robust. But, the migration methods so far are not able to eliminate background noise of GPR data.

Besides the migration and perception, GPR B-scan feature detection is also a significant topic since each B-scan feature defines subsurface target information. W. Al-Nuaimy et. al. from University of Liverpool \cite{al2000automatic} first proposed underground targets detection method by implementing Hough transform on GPR signals. They use back-propagation method to identify portions of the GPR image corresponding to target reflections. Moreover, as an automatic feature recognition method, \cite{cui2010automatic} applied center-surround difference detecting and fuzzy logic in GPR feature detection, which improve the tolerance of changes in viewpoint of GPR image. At last, as for deep learning method, \cite{lameri2017landmine} \cite{besaw2015deep} \cite{xu2018railway} \cite{pham2018buried} employed deep convolutional neural networks (DNNs) to extract meaningful signatures from 2D B-scan image and detect underground objects. However, towards data-driven visual inspection of subsurface targets reconstruction, there are still some challenges needed to be solved as following
\begin{itemize}
\item Firstly, the current GPR data collection still needs to be finished along either horizontal or perpendicular lines of a pre-defined survey grid triggered by GPR survey wheel, which is not robust and efficient for data collection.
\item Secondly, without the GPR data de-noised processing, background noise in GPR data would extremely effect migration result. 
\item Finally, a proper permittivity for subsurface material is highly relied on the pre-knowledge of geophysicist, there is still lack of a method to predict the dielectric based on GPR B-scan data.  
\end{itemize}

To address the above challenges, this paper developed an 3D GPR migration and object detection method to enable the localization and visualization of the subsurface targets. We first propose using visual inertial fusion to estimate the pose of the GPR device. GPR is triggered by survey wheel to collect A-scan data. After synchronizing the A-scan data with pose information, B-scan data is generated. Then, we employ a Faster R-CNN \cite{girshick2015fast} to locate the GPR B-scan hyperbolic features and output the corresponding bounding boxes. Once we got the Faster R-CNN detection results, we only keep the B-scan data in bounding boxes to do the migration while discard the rest of the data outside the bounding boxes, which has the effect to remove the noise. We propose the DepthNet to predict target depth. It takes the raw B-scan data with different dielectric values as input, and predicts the target depth in current GPR B-scan image. Finally, as depicted in Fig.\ref{fig:proposed_system}, our system performs migration algorithm based on the de-noised data, and reconstruct the target area based on our predicted depth of subsurface targets.

\section{System Architecture}

Our proposed system architecture is illustrated in Fig.\ref{fig:schmatic}, which is composed of three modules: 1) visual inertial fusion (VIF) module to generate pose information of GPR device,  2) DepthNet which detects B-scan of GPR image, extracts hyperbola features to remove the noise in B-scan data and predicts dielectric to determine the depth of the objects, 3) 3D GPR migration module which synchronizes the pose information with GPR scan data processed by DepthNet to reconstruct and visualize the 3D underground targets. Our goal is to enable the 3D subsurface object reconstruction and visualization, by means of our DepthNet based object detection and migration methods.

\subsection{Visual Inertial Odometry Positioning System}
In order to provide pose information for migration, we introduce visual inertial fusion (VIF) to obtain 6 DOF pose of GPR device. The VIF pipeline is illustrated in Fig.\ref{fig:schmatic}, where we use Intel D435i realsense camera which has Inertial Measurement Unit (IMU) and RGB-D camera embedded in the system. 

The VIF implementation is based on \cite{armesto2007fast}, where we fuse the IMU and the visual odometry in a loosely-approach, that is, the IMU performs pose estimation as the prediction while take visual odometry information as correction.

We first take the RGB and depth frames as input to initialize our pose and coordinate system. Meanwhile, IMU measurement is fed as propagation to estimate the pose of the IMU \cite{mourikis2007multi}. Finally, we fuse the pose calculated by visual odometry as the observation information to update the state. Therefore, the current camera pose $T_i \in \mathbb {SE}(3)$ could be estimated\cite{nutzi2011fusion}.

\subsection{GPR data collection and Labeling}
\label{subsection:gpr_dataset}
In order to facilitate the understanding of the raw data to help us to label the ground truth for our model training, we perform field data collection at a well designed test facility, i.e., Geophysical Survey Systems, Inc. (GSSI) test pit at Nashua, New Hampshire. For all collected data, it comes with the ground truth information including depth and length information of utility pipes, rebars and tanks. 

Our data collection follows the following steps which is highly recommended by the GSSI site engineers, 
\begin{itemize}
    \item Firstly, we set up a grid area with equal line spacing to cover the whole test pit.
    \item Secondly, we mount the camera on the GPR cart to generate pose, then we use ROS to subscribe  pose data and GPR scan data in order to make them synchronized.
    \item Finally, we move the platform in a zigzag pattern that defined in the first step and record both GPR sensory and pose information. A total of $140$ set of data were collected and each set contains an average of $800$ A-scan data synchronized with pose. 
\end{itemize}

Once we obtain all the field data, we label each single measurement in two aspects. First, this paper transform all the B-scan measurement into color image encoded with 'Hot' colormap. Then, we annotate the Region of Interest (ROI), that is, the hyperbolic feature with a bounding box which is described by $\{x_{min}, y_{min}, x_{max}, y_{max}\}$. Besides the bounding box, we also assign a dielectric value for each measurement as the ground truth. The system is used to predict dielectric is used to estimate the depth of the subsurface objects.

\begin{figure}[h]
    \centering
    \includegraphics[width=0.48\textwidth]{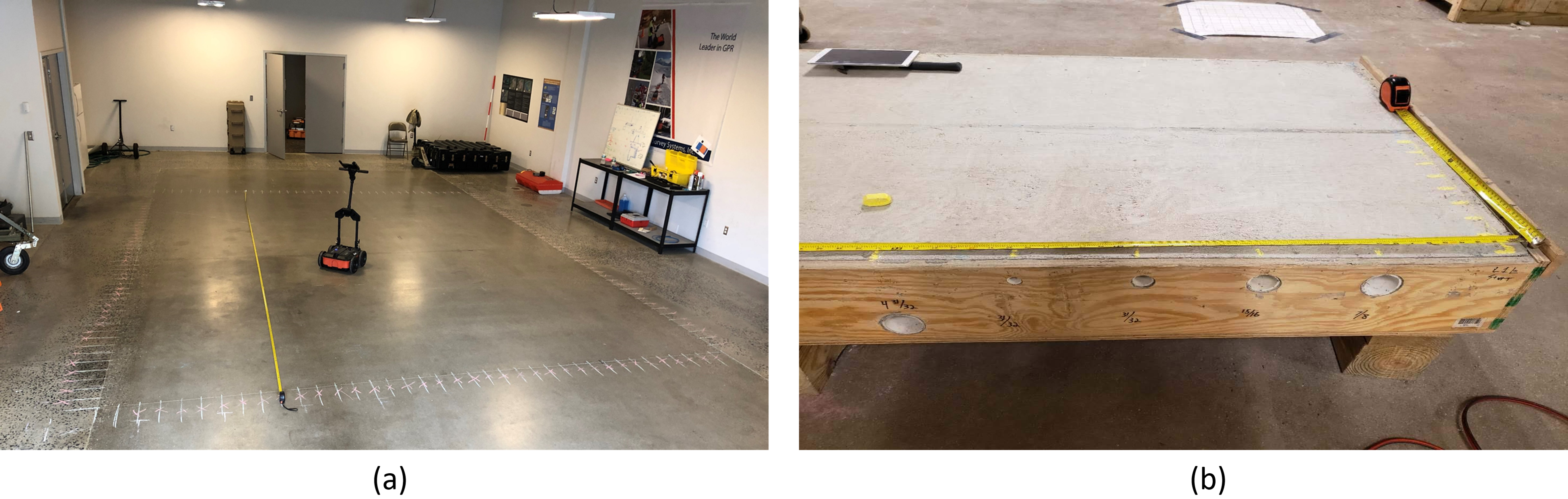}
    \caption{(a) Test pit in GSSI's garage, multiple targets buried underneath the surface. (b) Test slab in GSSI's slab room, multiple utility pipes inserted in the slab.}
    \label{fig:Data_Collection_GSSI}
\end{figure}

\subsection{GPR Migration Toward 3D Reconstruction}
\label{subsection: migartion}
Migration is a hyperbolic shape analysis approach to reconstruct the subsurface structure with 3D output, acting in a spatial deconvolution manner which belongs to Back Projection (BP) methodology \cite{demirci2012ground}. Migration is highly relied on dielectric of subsurface materials. It uses the dielectric value to calculate the signal propagation velocity in the medium from the hyperbolic features \cite{zhou2011fast}. Once the velocity is obtained, depth scale for subsurface targets could be reconstructed. 

Each GPR migration measurement, on a macroscopic scale, can be described by the well-known Maxwell’s equations (see Equ.\ref{equ:Maxwell}) \cite{jol2008ground}. The first order partial differential equations express the relations between the fundamental electromagnetic field quantities and their dependence on their sources \cite{giannopoulos2005modelling}.

\begin{equation}
\begin{aligned}
\label{equ:Maxwell}
    \nabla\cdot\vec{D} &= \frac{\rho}{\varepsilon_0} \\
    \nabla\cdot\vec{B} &= 0 \\
    \nabla\times\vec{E} &= -\frac{\partial \vec{B}}{\partial t} \\
    \nabla\times\vec{H} &= \mu_0\vec{J}+\mu_0\varepsilon_0\frac{\partial\vec{E}}{\partial t}
\end{aligned}
\end{equation}


Where $\vec{E}$ is the electric field strength vector (V/m); $\frac{\rho}{\varepsilon_0}$ is the electric charge density ($C/{m}^3$); $\vec{B}$ is the magnetic flux density vector (T) while $\vec{J}$ is the electric current ($A/{m}^2$); $\vec{D}$ is the electric displacement vector ($C/{m}^2$); $t$ is time (s) as well as $\vec{H}$ is the magnetic field intensity ($A/{m}$).

In Maxwell’s equations, the field vectors are assumed to be single-valued, bounded, and continuous functions of position and time. In order to simulate the GPR response from a particular target or set of targets, the above equations have to be solved subject to the geometry constraints of the problem and the initial conditions.

For migration, the first step is to calculate the two-way travel times (TWTT) of an electromagnetic wave from the transmitter to subsurface targets at which reflection occurs and returns back to receiver within the migration domain.

    \label{equ:xy}
    \begin{align}
    \mathcal{T}_{tr} &= 2 \cdot {D}_{tar}/v \\
    v &= {C} / \sqrt{\mathcal{D}}
    \end{align}    

where $\mathcal{D}$ represents different dielectric, which is the conductivity of the two materials (in this paper, we proposed a learning approach to obtain this parameter \ref{subsection:dielectric_traning}); ${C}$ denotes the velocity of light, ${D}_{tar}$ means depth of subsurface targets while $\mathcal{T}_{tr}$ means the two way travel time of GPR's antenna.

In second step, we implement the extrapolation of electromagnetic wave back to time domain by using two-dimensional Maxwell's equations.

\label{equ:Maxwell-two}
\begin{equation}
\begin{aligned}
    &\frac{\partial{E_{y}(x,z,t)}}{\partial{t}} = -\lambda E_y(x,z,t) + \frac{1}{\gamma(x,z)} \\ 
    &\quad\quad\quad\quad\quad\cdot [\frac{\partial{H_{x}(x,z,t)}}{\partial{z}} - \frac{\partial{H_{z}(x,z,t)}}{\partial{x}}]\\
    &\quad\quad\quad\quad\quad+ \frac{1}{\gamma{(x,z)}_{y}(x,z,t)}\\
    &\frac{\partial{H_{x}(x,z,t)}}{\partial{t}} = \lambda \frac {\partial{E_{y}(x,z,t)}}{\partial{Z}} \\
    &\frac{\partial{H_{z}(x,z,t)}}{\partial{t}} = \lambda \frac {\partial{E_{y}(x,z,t)}}{\partial{X}} \\
\end{aligned}
\end{equation}

where ${E_{y}(x,z,t)}$ is the electrical field in $y$ component. ${H_{x}(x,z,t)}$ and ${H_{z}(x,z,t)}$ are magnetic fields in ${x-}$ and ${z}$ components respectively. $\lambda$ represents the permittivity of the subsurface of material. $E^\prime_{y}(x,z,t)$ denotes the observed electric field in $y$ direction. In this way, the Maxwell's equations are solved by introducing the electromagnetic wave extrapolation, two-way travel times could be transferred into the distance between the GPR's antenna to the subsurface targets. 

The third step, we use Back Projection where the distance calculated in the second step is taken as the radius $r$. At each GPR measurement point, migration will take this point as the center and generate a semi-hemisphere with radius $r$. The potential target could be shown up on any points located at the surface of this semi-hemisphere. Along with the movement of GPR measurement, there will be more semi-hemispheres with different radius get generated, their intersection should be the location of the targets. By this way, a 3D migration image could be generated.

Finally, we propose a new method on GPR migration. The traditional way can only achieve pseudo 3D GPR imaging because the GPR data is collected along either horizontal or perpendicular directions of the pre-defined survey grid. However, by receiving rotation information from VIF pose estimation system, the GPR cart is able to collect the GPR data in any directions, thus constructs real 3D GPR image. In Equ.\ref{equ:antenna}, $A\_{pre}$ denotes the coordinate of previous antenna while $A\_{update}$ represents the coordinate of updated antenna. $\theta$ is the antenna rotation angle.

\begin{equation}
    {\left[\begin{array}{c}
    A_{update\_{x}} \\
    A_{update\_{y}} \\
    \end{array}
    \right]}={
    \left[\begin{array}{cc}
    \cos\theta & -\sin\theta \\
    \sin\theta & \cos\theta \\
    \end{array}
    \right]}\times 
    {\left[\begin{array}{c}
    A_{pre\_{x}} \\
    A_{pre\_{y}} \\
    \end{array}
    \right]}
    \label{equ:antenna}
\end{equation}

\subsection{DNN based Target Detection and Depth Prediction}
\label{subsection: detected_migartion}

Migration is a process that transforms the 2D B-scan into the 3D GPR imaging. However, due to subsurface noise and the uncertainty of the material dielectric, it is almost impossible to reconstruct the real 3D GPR imaging accurately. Thus, we propose two models to solve the problems, which are GPR object detection model and dielectric prediction model. The details of the DNN network is illustrated in Fig.\ref{fig:schmatic}, where we use Faster R-CNN \cite{szegedy2017inception} to perform subsurface object detection and a new DepthNet to predict the dielectric of the subsurface material.


\subsubsection{GPR Based Object Detection}

In order to produce a clear 3D GPR imaging from 2D B-scan data, we use Faster R-CNN \cite{szegedy2017inception}, an inception architecture with residual connections network, to detect B-scan hyperbolic features. Then, we take the detected bounding boxes as the region of interest (RoI). The data outside the RoI are considered as noise which is expected to be discarded in migration process. Moreover, more object detection networks are used in this section in order for comparison of Intersect of Union (IoU). \cite{huang2017speed}.

\begin{figure}
    \centering
    \includegraphics[width=0.485\textwidth]{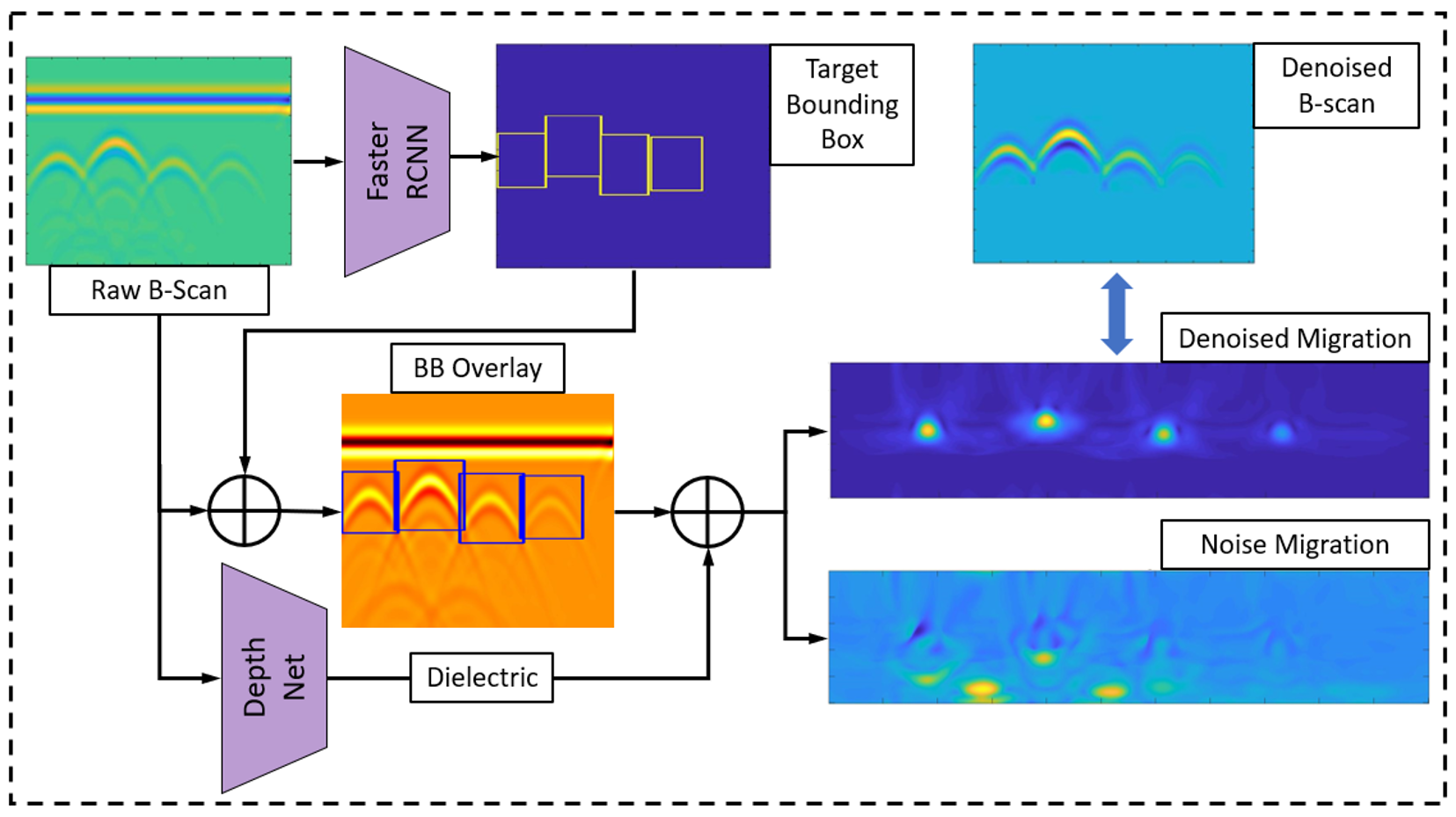}
    \caption{Proposed DepthNet framework. DepthNet consists of two parts: (1) B-scan Detection and (2) Noise Cancellation.}
    \label{fig:3D_migration_visualization}
\end{figure}

\subsubsection{DepthNet for Depth Prediction}

In our work, the DepthNet outputs the $d_{t(i)}, i \in \{0,1,...\}$ for each of embedded targets. As illustrated in Fig.\ref{fig:schmatic}, the DepthNet consists of three sub-nets, where the lowermost model predicts the dielectric of the subsurface material, the uppermost and the middle model are fully-connected nets which take the Faster R-CNN bounding boxes and features as the input. The dielectric model takes the raw B-scan image as input and resize it to $224 \times 224$ while the encoder is Resnet$\-$101. Then DepthNet takes the dielectric model output features and the other two layers features as the input to predict the depth of each target. 

\textbf{Loss Design:} After predicting the depth of each target and the dielectric, we optimize the model by using a weighted sum error approach to regress the model, 

\begin{equation}
    \centering
    L_{loss} = \lambda _{\mathcal{D}}\frac{\sum_{i=0}^{n-1} (^\mathcal{D}y_{i} - ^\mathcal{D}y_{i}^{p})}{n} + \lambda _{D} \frac{\sum_{i=0}^{n-1} (^{D}y_{i} - ^{D}y_{i}^{p})}{n}
    \label{equ:faster_rcnn}
\end{equation}

where $^{D/\mathcal{D}}y_{i}$ is the ground truth depth, $^{D/\mathcal{D}}y_{i}^{p}$ is the depth prediction, $\lambda _{\mathcal{D}}$ and $\lambda _{D}$ are the weight for dielectric and depth loss respectively. In this paper, we employ mean square error loss to regress the depth prediction model.

\subsection{Target Visualization}

In order to visualize and reconstruct the subsurface objects, we combine DNN based GPR images with our proposed migration method. As it is illustrated in Fig.\ref{fig:3D_migration_visualization}, we considered detected B-scan hyperbolic features and the corresponding bounding boxes as the region of interest (RoI). Then, only the GPR B-scan data in RoI would be fed into migration processing in order to obtain the de-noised subsurface objects. In this way, we can easily estimate and localize the subsurface objects as well as the depth information for each targets, where the depth information is what also obtained from DepthNet. Besides, the noise map we recovered from the removed noise is also considered as an evaluation of our proposed system.

\begin{figure*}
    \centering
    \includegraphics[width=0.9\textwidth]{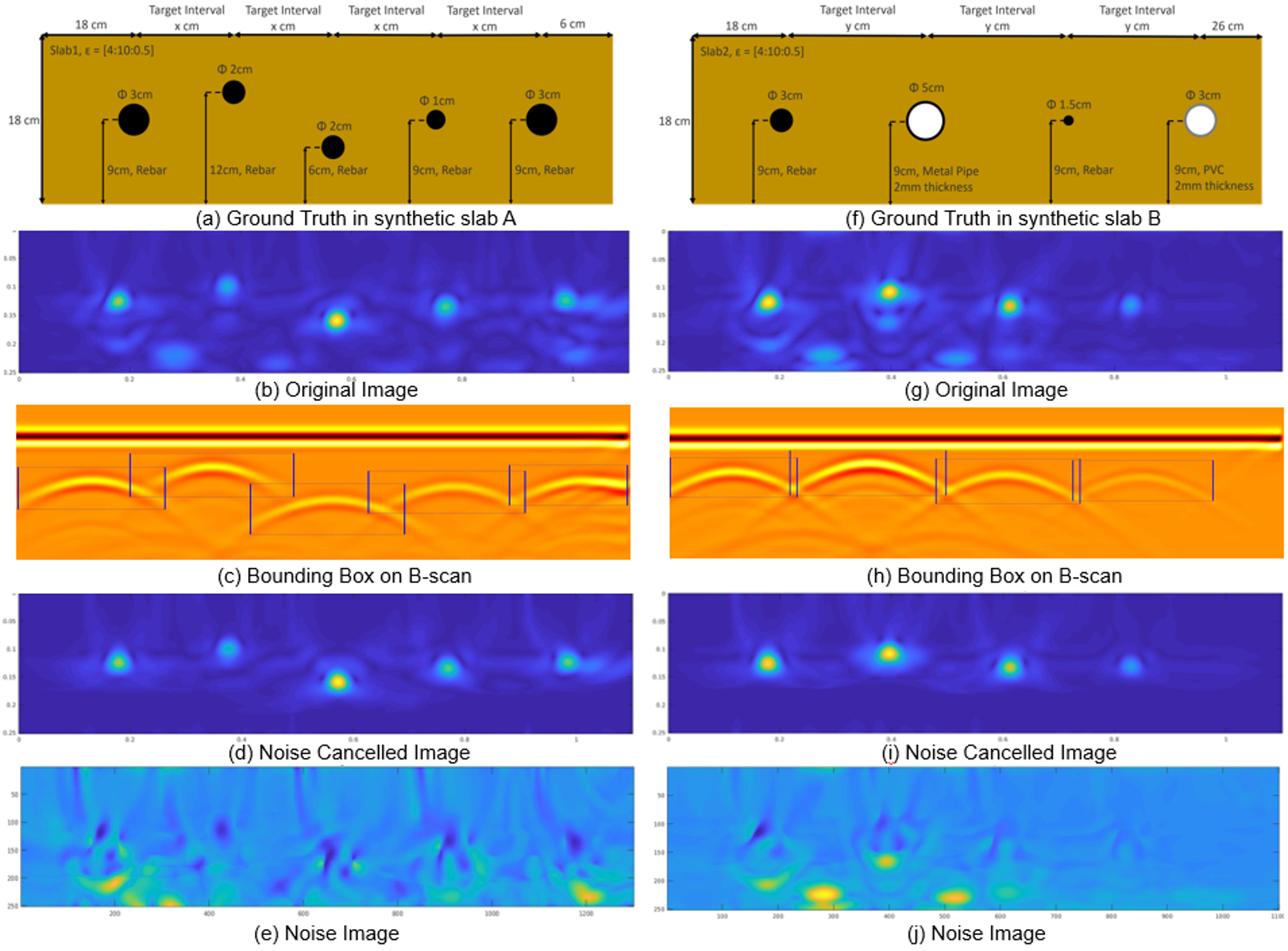}
    \caption{Front view of fine tune migration method compared with traditional migration method.}
    \label{fig:slab_compare}
\end{figure*}

\begin{figure}
    \centering
    \includegraphics[width=0.485\textwidth]{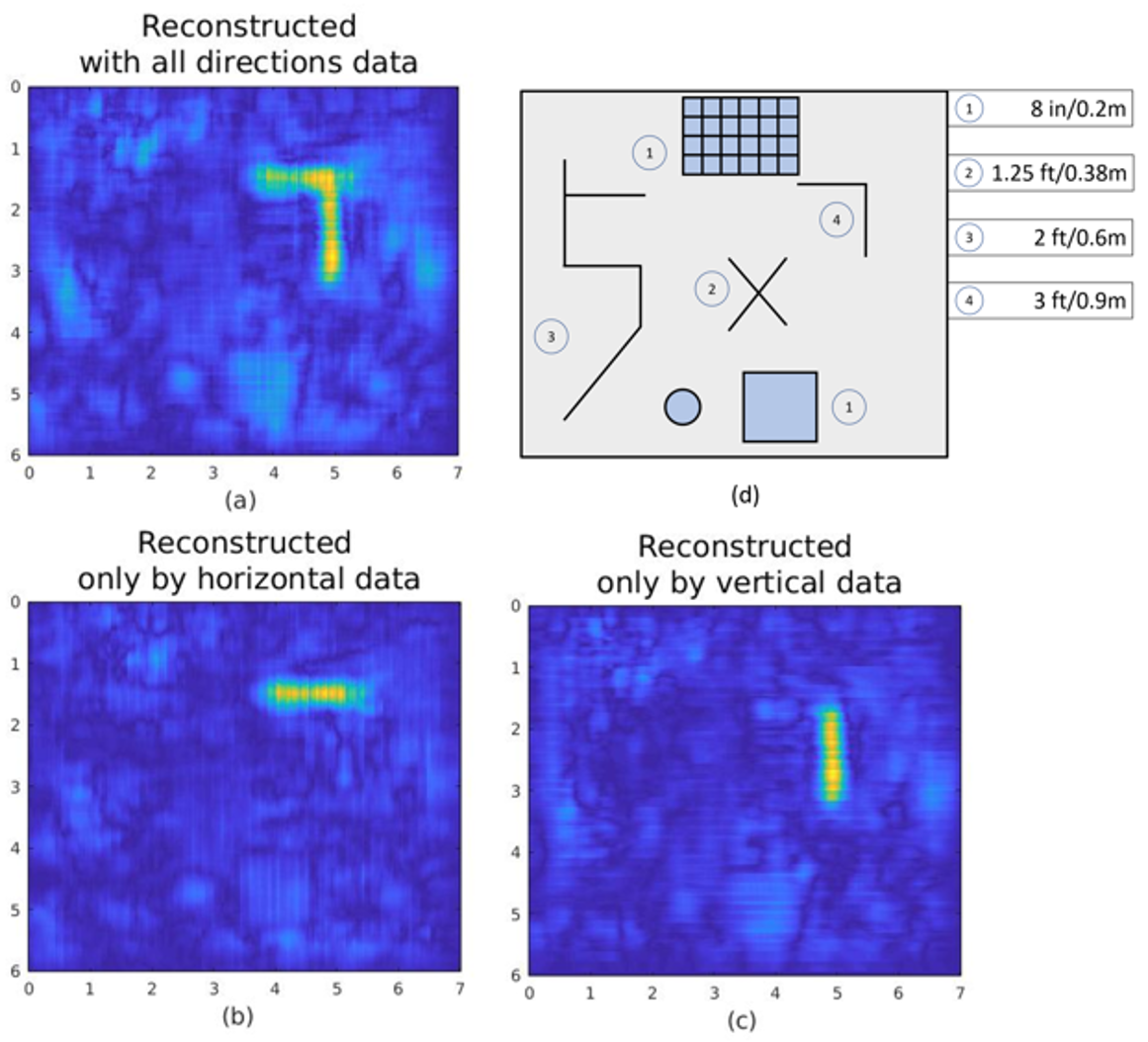}
    \caption{Proposed migration results in GSSI's test pit compare with ground truth and traditional migration methods. (a) shows the target reconstruction result used by proposed migration method; (b) and (c) shows the limitation of traditional method; (d) shows the ground truth of subsurface targets, which are located at different layers with different depth. All the graphs show above are in top view. }
    \label{fig:Migration}
\end{figure}

\section{Experiments}

We verify our GPR underground objects reconstruction system on the dataset we created according to Section \ref{subsection:gpr_dataset}. We illustrate our newly proposed migration algorithm and demonstrate the effectiveness the 3D GPR subsurface reconstruction by using our learning based approach. For all these evaluations, they are conducted on an GPU server, with Intel Core i9-9900K 3.2GHz CPU, GeForce RTX 2080 Ti GPU, and 32GB RAM.

\subsection{Performance of Migration}
\label{subsection: migration_experiment}
In order to verify the migration method we proposed in this paper, a 3D subsurface image of test pit is first generated and compared with the ground truth provided by GSSI. The pose information obtained from VIF estimation system and the GPR scan data are considered as the input for migration. The target in the Fig.\ref{fig:Migration} is a right angle shape pipe, which located at 0.9 meter beneath the surface of test pit. By implementing the traditional method, targets could only be reconstructed either in horizontal direction or vertical direction, which will make the migration result be not intact. However, different dielectric value could also influence the accuracy of migration. Due to this reason, the necessary of the depth prediction of underground targets will be testified in the next section.

\subsection{Performance Comparison of GPR Target Detection and Depth Prediction}
\label{subsection:dielectric_traning}
\textbf{Object Detection Comparison}

Table.\ref{table:network_compare} shows the results of B-scan feature detection compared with different models \cite{pham2018buried,szegedy2017inception,liu2016ssd, redmon2018yolov3}. The results show that Faster R-CNN resnet101 has the best performance on GPR based object detection. Once we obtain the detection results, our proposed DepthNet is implemented as the dielectric prediction, that is actually, subsurface targets depth prediction. In Table.\ref{table:network_compare}, mAP@IoU = 0.75 is mean average precision at 75\% IoU, mAP@IoU = 0.50 is mean average precision at 50\% IoU, AR@10 is average recall with 10 detections, AR@100 is average recall with 100 detections.

\begin{table}
\caption{Detection Performance Comparison.}
\label{table:network_compare}
\begin{center}
\begin{tabular}{|c|c|c|c|c|}
\hline
\hline
&\multicolumn{2}{|c|}{mAP@IoU} &\multicolumn{2}{|c|}{AR}\\
\cline{2-5}
Models&\multicolumn{1}{|c|}{$0.75$} &\multicolumn{1}{|c|}{$0.50$} &\multicolumn{1}{|c|}{$10$} &\multicolumn{1}{|c|}{$100$}\\
\hline
\hline
YOLO v3 & $89.6$ & $89.3$ & $90.2$ & $91.2$\\
ssd mobilenet v1 & $85.8$ & $88.6$ & $84.2$ & $84.9$\\
ssd mobilenet v2 & $86.1$ & $87.8$ & $87.7$ & $87.0$\\
ssd Inception v2 & $88.4$ & \textbf{90.1} & $88.9$ & $85.6$\\
ssdlite mobilenet v2 & $82.2$ & $83.3$ & $89.2$ & $89.3$\\
Faster R-CNN Inception v2& $89.3$ & $89.3$ & $92.6$ & $91.6$\\
Faster R-CNN resnet101 & \textbf{90.5} & $89.0$ & \textbf{92.2} & \textbf{92.2}\\
Faster R-CNN resnet50 & $89.6$ & {89.8} & $90.8$ & $91.9$\\
\hline
\end{tabular}
\end{center}
\end{table}

\textbf{Network Training}


\begin{figure}[h]
    \centering
    \includegraphics[width=0.4\textwidth]{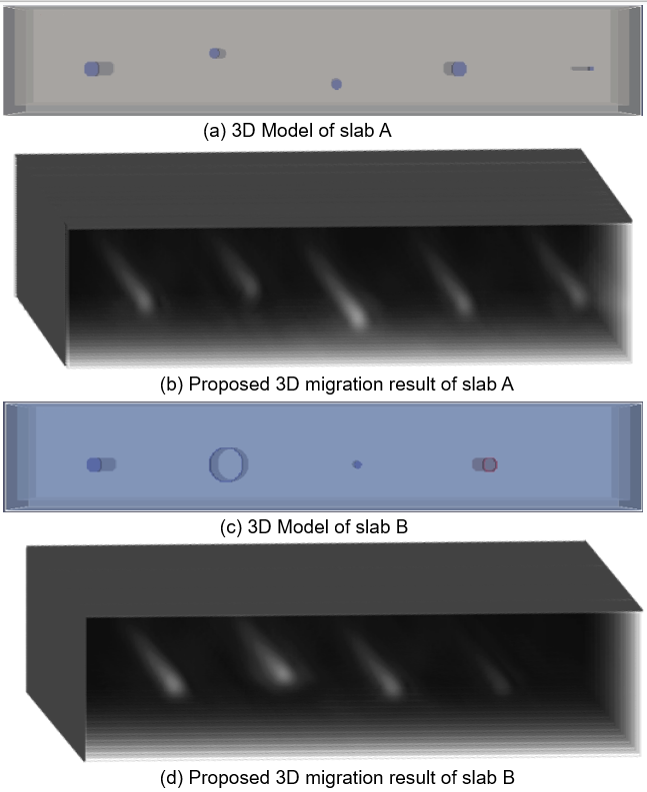}
    \caption{3D target migration map compared with the original 3D target model. (a), (c) are the 3D model of synthetic slabs while (b), (d) show the proposed 3D target migration results.}
    \label{fig:highlight_3d_model}
\end{figure}

We trained DepthNet on a RTX 2080Ti GPU server and used Pytorch to deploy the algorithm. For the DepthNet, we used the stochastic gradient decent (SGD), with an initial learning rate at $4e-5$, momentum as $0.9$, and weight decay of $5e{-5}$. The model loss at last converged to $0.0524$

\textbf{Accuracy Evaluation}

We perform the dielectric prediction validation using the test data from the GPR dataset we created in Section.\ref{subsection:gpr_dataset}. For testing purpose, we use total of $50$ B-scan images which have the different dielectric. The accuracy of the model is tested from two different aspects: 1) the individual accuracy for B-scan feature dielectric prediction; 2) the average accuracy of the depth prediction which is $0.112$.

\textbf{Visualization for Proposed Method}

To visualize the results of the GPR based feature detection, which used as noise cancellation for GPR migration, this paper first shows the ground truth of the two different synthetic slabs created by gprMax. Then, in each B-scan raw data, we overlays the detected bounding boxes on it. Besides, the predicted dielectric also be registered in migration as the pre-processing for 3D subsurface targets reconstruction. Then, we compared the original migration result and noise cancelled migration results in front view, since we could also compare these results with the bounding boxes overlapped B-scan data. At last, the cancelled noise images are also attached in order to validate our proposed method. In Fig.\ref{fig:slab_compare} (a), (f) shows the ground truth of the two synthetic slabs generated by gprMax. This two slabs are embedded with conductive and dielectric rebar, steel pipe, another rebar and PVC pipe with different size from left side to right. Then (b), (c) and (g), (h) compares the migration results before implementing B-scan feature detection and after detection mentioned in \ref{subsection: detected_migartion}. At last, (e), (j) show the noise we cancelled with proposed method.

\subsection{3D Object Migration Map}

This section uses the results from the previous two sections to generate the 3D object migration map. First, our proposed migration method provide the ability for real 3D subsurface targets reconstruction. Then, by detecting the B-scan features, the noise in B-scan raw data could be removed. Moreover, our proposed DepthNet could provide the depth information based on the input B-scan raw data. By combining these three methods, Fig.\ref{fig:highlight_3d_model} shows the final comparison results of the 3D targets reconstruction.


\section{CONCLUSIONS}
This paper introduces an DNN based 3D GPR imaging system, which is able to locate and visualize the subsurface objects. First, this system implements visual inertial fusion to estimate the pose of the GPR sensor. Then, we propose an improved random motion migration method which eliminates the limitation of current GPR data collection procedure which requires the straight line motion along survey grid. After that, DNN based target detection is employed, by only processing the B-scan data in detected bounding boxes, background noise in raw B-scan image could be removed. Finally, the proposed DepthNet is used to predict the depth of subsurface objects, according to the estimation of dielectric characteristic of the material. The experiments show the effectiveness of our proposed 3D subsurface objects reconstruction methodology.

\section{Acknowledgement}

Research was supported in part by NSF grant number IIP-1915721. J Xiao has significant financial interest in InnovBot LLC, a company involved in R\&D and commercialization of the technology. 


\bibliographystyle{IEEEtran}  
\bibliography{icra2019}

\end{document}